\documentclass[runningheads]{llncs}

\newif\ifdraft
\draftfalse

\usepackage{graphicx}
\usepackage{color}
\usepackage{tabularx}
\usepackage{amsmath}
\usepackage{subcaption}
\usepackage{centernot}
\usepackage{placeins}
\usepackage{comment}
\usepackage[misc]{ifsym}
\usepackage[hyphens]{url}
\usepackage[multiple]{footmisc}

\newcommand{\afabcomment}[1]{\ifdraft{\leavevmode\color{green}{[AF]: {#1}}}\else{\vspace{0ex}}\fi}
\newcommand{\gascomment}[1]{\ifdraft{\leavevmode\color{blue}{[GAS]: {#1}}}\else{\vspace{0ex}}\fi}
\newcommand{\gmdncomment}[1]{\ifdraft{\leavevmode\color{red}{[GMDN]: {#1}}}\else{\vspace{0ex}}\fi}

\begin{document}
\title{Incentives for Item Duplication under Fair Ranking Policies}
 \author{Giorgio Maria Di Nunzio \and Alessandro Fabris \textsuperscript{\Letter}  \and Gianmaria Silvello \and  Gian Antonio Susto}

%\author{First Author\inst{1}\orcidID{0000-1111-2222-3333} \and
%Second Author\inst{2,3}\orcidID{1111-2222-3333-4444} \and
%Third Author\inst{3}\orcidID{2222--3333-4444-5555}}

%
\authorrunning{G. Di Nunzio et al.}

%\institute{Princeton University, Princeton NJ 08544, USA \and
%Springer Heidelberg, Tiergartenstr. 17, 69121 Heidelberg, Germany

\institute{University of Padua, Padua, Italy \\
\email{\{dinunzio, fabrisal, silvello, sustogia\}@dei.unipd.it}
}

\maketitle              % typeset the header of the contribution
\begin{abstract}
Ranking is a fundamental operation in information access systems, to filter information and direct user attention towards items deemed most relevant to them. Due to position bias, items of similar relevance may receive significantly different exposure, raising fairness concerns for item providers and motivating recent research into fair ranking. While the area has progressed dramatically over recent years, no study to date has investigated the potential problem posed by duplicated items. Duplicates and near-duplicates are common in several domains, including marketplaces and document collections available to search engines. In this work, we study the behaviour of different fair ranking policies in the presence of duplicates, quantifying the extra-exposure gained by redundant items. We find that fairness-aware ranking policies may conflict with diversity, due to their potential to incentivize duplication more than policies solely focused on relevance. This fact poses a problem for system owners who, as a result of this incentive, may have to deal with increased redundancy, which is at odds with user satisfaction. Finally, we argue that this aspect represents a blind spot in the normative reasoning underlying common fair ranking metrics, as rewarding providers who duplicate their items with increased exposure seems unfair for the remaining providers.

\end{abstract}

\section{Introduction}
%\afabcomment{The necessity of ranking, the chance for fairness over repeated impressions, the interplay of fairness and duplication. RQs and summary of answers.}

%\afabcomment{Individual fairness is attractive due to subsuming group fairness, however it may be gameable.}

%\afabcomment{Monopoly}

Ranking is a central component in search engines, two-sided markets, recommender and match-making systems. These platforms act as intermediaries between providers and consumers of items of diverse nature, facilitating access to information, entertainment, accommodation, products, services, jobs and workers. 
%a visualization necessity in online systems. More in general, when a set of items is presented to a human, an order of presentation is inevitably defined. 
The rank of an item in a result page is a strong predictor of the attention it will receive, as users devote most of their attention to the top positions in a list, and are less likely to view low-ranking items \cite{joachims2007:se}. This \emph{position bias} is at the root of fairness concerns for providers of ranked items, as comparably relevant results may receive remarkably different exposure. 
%This is especially true when
%attention is strongly skewed towards the top of rankings.
%, as the top position lends strong credibility to an item,
%; as a result, much user attention is funneled to it. 
In the absence of countermeasures, unfair exposure can affect item providers on e-commerce websites, job-search platforms and commercial search engines, such as Amazon sellers, Airbnb hosts, job candidates on LinkedIn and owners of contents ranked by Google \cite{biega2018:ea,geyik2019:fa}.
%Examples of famous platforms where the impact of unfair exposure on item providers is immediately tangible are, among many others, Amazon, Airbnb and Google.

Unfair exposure can compound and increase over time, as the same query, issued multiple times to a system, is met with the same ranking. Each time the query is processed by the system, items gain a fixed, potentially unfair, level of exposure; this is a severe problem with static ranking policies, which map relevance scores to rankings in a deterministic fashion. Non-static policies, on the other hand, can respond to identical queries with different rankings, and they are more suited to equitably distribute exposure among relevant items. In recent years, fair ranking policies and measures have been proposed, which consider repetitions of the same query and encourage rotation of relevant items in the top-ranked positions \cite{biega2019:ot,biega2018:ea,diaz2020:es,singh2018:fe,thonet2020:mg}. 

\begin{figure}[t]
\centering
\includegraphics[width=0.8\textwidth]{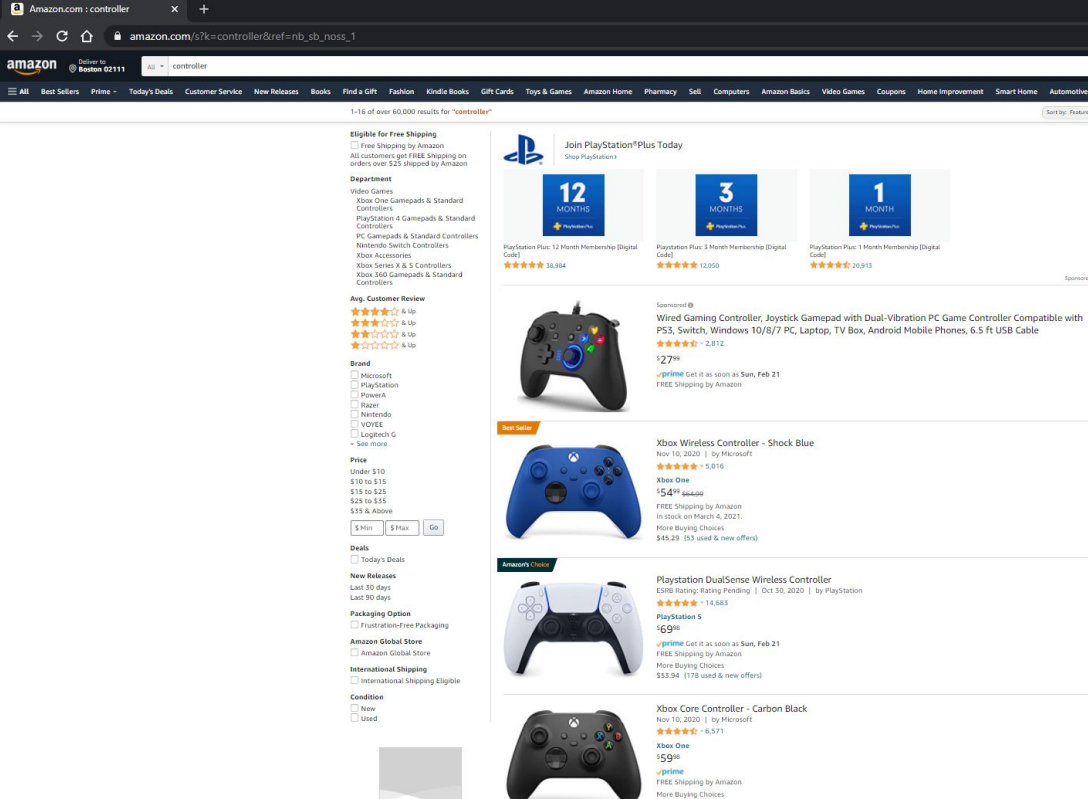}
\caption{Amazon result page for query \texttt{controller} issued on February 19, 2021 by a Boston-based unregistered user in \emph{incognito} browser mode. Top 4 results comprise near-duplicates in positions 1 and 3 (0-based indexing).} \label{fig:dupl_amaz}
\end{figure}

While these measures and approaches are surely a sign of solid progress in the area of fair ranking, in this paper we highlight a potential blind spot in their normative reasoning: duplicates. Item duplicates and near-duplicates are not uncommon in online domains such as e-commerce websites \cite{amazon_split} and online document collections \cite{frobe2020:ec}. Anecdotal evidence for this phenomenon can be found in official forums for item providers of popular marketplaces\footnote{\url{https://community.withairbnb.com/t5/Hosting/Unfair-duplication-of-same-listing-to-gain-more-exposure/td-p/850319}, all links accessed on 02-03-21}\footnote{\url{https://community.withairbnb.com/t5/Help/Duplicate-photos-in-listings-and-terms-of-service/td-p/1081009}}\footnote{\url{https://sellercentral.amazon.com/forums/t/duplicate-search-results/445552}} and in Figure \ref{fig:dupl_amaz}.
Based on the reasoning brought forth in recent works, requiring equal exposure for equally relevant items \cite{biega2018:ea,diaz2020:es}, two copies of the same item deserve more exposure (in sum) than a single copy. On the contrary, in some situations it is reasonable to postulate that multiple copies of an item deserve the same attention the item would be entitled to on its own, especially if the copies benefit the same provider. 

The key contribution of this work is to analyze the tension between fairness and diversity when duplicates are not properly considered and factored into the fair ranking objective. More in detail, we show that, under different settings, common fair ranking policies reward duplicates more than static policies solely focused on relevance. We argue that this phenomenon is unfair to the providers of unique (non-duplicate) items and problematic for system owners and users as it introduces an incentive for redundancy. The rest of this paper is organized as follows. Section \ref{sec:rel} introduces related work, covering fairness and diversity in ranking. Section \ref{sec:central} is the core of our study, where we formalize the problem and analyze the benefits obtained by providers through item duplication. We quantify the extra-attention earnt by duplicated items in a controlled setting, as we vary the relevance of the items available to a system, its ranking policy and the cost of duplication. Section \ref{sec:concl} contains closing remarks and outlines directions for future work.

\section{Related Work}
\label{sec:rel}
\subsection{Fairness in Ranking}
\label{sec:fair}
%\afabcomment{Motivation. Cross-query or within query measures, stateful or stateless approaches.}

Fairness in ranking requires that the items ranked by a system receive a suitable share of exposure, so that the overall allocation of user attention is considered fair according to a criterion of choice \cite{biega2018:ea,singh2018:fe}. Fair ranking criteria depend on the specific context and normative reasoning, often inheriting and adapting notions from the machine learning fairness literature, such as independence and separation \cite{barocas2019:fm}.
%Several measures have been proposed, concentrating both on group fairness \cite{yang2017:mf} and individual fairness \cite{diaz2020:es}.

\emph{Position bias} and \emph{task repetition} are peculiar aspects of many fair ranking problems. \emph{Position bias} refers to the propensity of users of ranking systems to concentrate on the first positions in a list of ranked items, while devoting less attention to search results presented in lower positions \cite{joachims2007:se}. Common measures of accuracy in ranking, such as Expected Reciprocal Rank (ERR) \cite{chapelle2009:er}, hinge on this property: they reward rankings where items are presented in decreasing order of relevance, so that the positions which attract most user attention are occupied by the most relevant items. These are static ranking measures, which summarize the performance of a system with respect to an information need by modeling a single user-system interaction. However, users can issue the same query multiple times, requiring a search engine to repeatedly attend to the same task (\emph{task repetition}). Repeated queries, stemming from the same information need, are sometimes called \emph{query impressions}.\footnote{\url{https://fair-trec.github.io/2020/doc/guidelines-2020.pdf}}

Recently, several measures of fairness in rankings have been proposed, which take into account the peculiarities of ranking problems \cite{biega2018:ea,diaz2020:es,singh2018:fe}. These measures incorporate \emph{position bias}, by suitably modeling user browsing behaviour when estimating item exposure, and consider \emph{task repetition} by evaluating systems over multiple impressions of the same query, thus encouraging rotation of relevant items in top ranks. For example, \emph{equity of amortized fairness} \cite{biega2018:ea} considers cumulative attention and relevance of items over multiple query repetitions, and is defined as follows. ``A sequence of rankings $\rho^1, \dots, \rho^J$ offers equity of amortized attention if each subject receives cumulative attention proportional to her cumulative relevance'', where the accumulation and amortization process are intended over multiple queries and impressions.

Depending on its amortization policy, measures of fairness in rankings can be (i) \emph{cross-query} or (ii) \emph{within-query}. (i) Cross-query measures are aimed at matching cumulative attention and relevance across different information needs \cite{biega2018:ea}; this approach has the advantage of naturally weighing information needs based on their frequency and to enforce fairness over a realistic query load. On the downside, these fairness measures may end up rewarding systems that display irrelevant items in high-ranking positions. (ii) Within-query measures, on the other hand, enforce fairness over impressions of the same query \cite{diaz2020:es}; this amortization policy results in one measurement for each information need and does not run the risk of rewarding systems that compensate an item's exposure across different information needs, which may result in balancing false negatives (missed exposure when relevant) with false positives (undue exposure when irrelevant).

Different approaches have been proposed to optimize ranking systems against a given fairness measure. Most of them make use of the \emph{task repetition} property by employing stochastic ranking policies. These systems are non-deterministic since, given a set of estimated relevance scores, the resulting rankings are not necessarily fixed. A key advantage of stochastic ranking policies over deterministic ones lies in the finer granularity with which they can distribute exposure over multiple impressions of a query. Depending on whether they keep track of the unfairness accumulated by items, policies can be \emph{stateless} or \emph{stateful}. Stateless systems are based on drawing rankings from an ideal distribution independently \cite{diaz2020:es,singh2018:fe,singh2019:pl}. This family of approaches can yield high-variance exposure for items, especially over few impressions, due to rankings being independent from one another. Moreover, they are not suitable to target cross-query measures as they would require estimating the future query load. Stateful solutions, on the other hand, keep track of the unfairness accumulated by items \cite{biega2018:ea,morik2020:cf,thonet2020:mg}, and exploit it to build controllers or heuristics that can actively drive rankings toward solutions that increase the average cumulative fairness. 
%On the downside, when targeting within-query measures these approaches need a large volume of bookkeeping, increasing with the number of distinct information needs satisfied by the ranking system.

\subsection{Diversity in Ranking}

\gmdncomment{Questa sezione è potenzialmente sbilanciata rispetto a quella sopra.} \gascomment{Vero, ma le terrei comunque distinte.}

The diversity of items in search results is important for users of products and services that feature a ranking component  \cite{broder2000:if,clarke2008:nd,ekstrand2014:up}. In the absence of ad-hoc approaches measuring and favouring diversity, very similar items may be present in a result page \cite{broder2000:if,nishimura2019:il}. Duplicates and near-duplicates are the most severe example of redundant results \cite{bernstein2005:rd,clarke2008:nd}. 

Redundant items are present in collections for web search \cite{bernstein2005:rd}, with repercussions on search engines that need to handle duplicates at different stages of their life cycle, including training, testing and deployment \cite{broder2000:if,frobe2020:sb,frobe2020:ec}. Moreover, duplicate or near-duplicate listings of items can be present in online marketplaces, such as Airbnb and Amazon \cite{amazon_split}. Multiple listings for the same item can derive from a legitimate need to highlight slightly different conditions under which a product or service is provided, or from an adversarial attempt to increase visibility on the platform through redundancy. Especially in the latter case, duplicates are viewed unfavourably by system owners for their negative impact on user experience \cite{amazon}.

\section{Duplicates and Fair Ranking}
\label{sec:central}
In this section, we illustrate the mechanism through which fairness in ranking may reward duplicates by granting multiple copies of the same item more exposure than it would obtain as a single item. Section \ref{sec:rew} introduces the problem, explains its root cause and presents a basic model of duplication and its cost. Section \ref{sec:exp} details the synthetic experimental setup, summarizing the key parameters of the problem and the values chosen for the present analysis. Results are reported in Section \ref{sec:res}. The relevant notation is summarized in Table \ref{tab:notation}.

\begin{table}
\centering
\caption{Notation employed in this work. \gascomment{Forse anticipata su  visto che rho lo usiamo prima?}}\label{tab:notation}
\begin{tabular}{ll}
\hline
Symbol & Meaning \\
\hline
$u_i$ &  items to be ranked by system, $i \in \left \lbrace 1, \dots, I \right \rbrace$\\
$u_{\tilde{i}}$ & duplicate of item $u_i$ \\
$q_j$ &  query impressions issued to system, $j \in \left \lbrace 1, \dots, J \right \rbrace$\\
$\rho^j$ &  ranking of items in response to query $q_j$\\
$\rho_i^j$ &  rank of item $u_i$ in ranking $\rho^j$\\
$\pi$ & a ranking policy \\
$\rho_{\pi}$ & $\left \lbrace \rho^1, \dots, \rho^J \right \rbrace$ sequence of rankings obtained via policy $\pi$ \\
$u(\rho_{\pi})$ & utility function rewarding ranking sequence based on user satisfaction \\
$a_i^j$ & attention received by item $u_i$ in ranking $\rho^j$\\
$r_i^j$ & relevance of item $u_i$ for the information need expressed by $q_j$ \\
$k$ & cost of duplication, such that $r_{\tilde{i}}^j=k r_i^j, \hspace{0.2cm} k \in (0,1)$ \\
$\delta_{i+1,i}^j$ & difference in relevance for adjecently ranked items (simplified to $\delta$) \\
$A_i$ & $\sum_{j=1}^J a_i^j$, i.e.\ cumulative attention received by item $u_i$ \\
$R_i$ & $\sum_{j=1}^J r_i^j$, i.e.\ cumulative relevance of item $u_i$ over queries $\left \lbrace q_1, \dots, q_J \right \rbrace$\\
$f(A, R)$ & fairness function, defining the ideal relationship between $A_i$ and $R_i$ \\
\hline
\end{tabular}
\end{table}

\subsection{Rewarding Duplicates}
\label{sec:rew}
%\afabcomment{Brief explainer on how redundancy may be favoured. Our set up (incl. cost of duplication).}

%In the absence of ad-hoc approaches measuring and favouring diversity, very similar items may be heavily featured in result pages \cite{broder2000:if,nishimura2019:il}, with a negative impact on user satisfaction \cite{ekstrand2014:up}. Duplicates and near duplicates are an extreme version of this redundancy problem \cite{clarke2008:nd,bernstein2005:rd}. 

If a ranking approach is purely based on the relevance of single items and unaware of their potential redundancy, two copies of the same item will receive more attention than a single copy. For example, let us consider an item $u_i$ ranked in position $\rho_i^j=n$, receiving in turn a share of user attention $a_i^j$. If a copy $u_{\tilde{i}}$ is created, in the absence of a ranking signal that rewards diversity, the item pair will rank in positions $\rho_i^j=n$, $\rho_{\tilde{i}}^j=n+1$ (or viceversa). As a result, under a non-singular attention model, the sum of attentions received by the item pair is greater than the original $a_i^j$.

The above consideration holds true regardless of notions of fairness in rankings. However, in the presence of fairness constraints, there may be a further advantage for duplicates. For example, under \emph{equity of amortized fairness} \cite{biega2018:ea}, which requires cumulative exposure of items proportional to their cumulative relevance, two items of identical relevance deserve twice as much exposure as a single item. 

%In terms of normative reasoning for fairness in the presence of duplicates, it is important to consider the entities and subjects who benefit from exposure of an item. If a subject creates 

In reality, there are several factors that make duplication ``expensive'' in terms of the features commonly exploited by systems for item retrieval and ranking. Firstly, some of these features, such as user ratings, stem from the interaction of users with items; if an item is duplicated, its interactions with users will be distributed across its copies, presumably reducing their relevance score and lowering their rank in result pages. Moreover, a ranking system may explicitly measure the diversity of retrieved items \cite{clarke2008:nd} and favour rankings with low redundancy accordingly \cite{nishimura2019:il}. Finally, some platforms forbid duplication, in the interest of user experience, and enforce this ban with algorithms for duplicate detection and suppression procedures \cite{amazon}.

Therefore, we treat duplication as an expensive procedure, with a negative impact on items' relevance scores. We assume that the cost of duplicating an item only affects the new copy, while leaving the relevance of the original copy intact. We model duplication cost as a multiplicative factor $k \in (0,1)$, reducing the relevance score of new copies of an item. In other words, if a copy $u_{\tilde{i}}$ of item $u_i$ is created, then $r_i$ remains constant, while $r_{\tilde{i}}=kr_i$. Richer models of duplication cost are surely possible (e.g.\ also reducing the relevance $r_i$ of the original copy), and should be specialized depending on the application at hand. 

\subsection{Experimental Setup}
\label{sec:exp}
%\afabcomment{Our optimization problem (incl. attention model, utility of ranking (nERR) and fairness function). Approaches (PL, greedy stateful). Relevance profiles. Costs of duplication.}

%Fair ranking can be cast as a constrained optimization problem looking to optimize fairness with relevance constraints \cite{biega2018:ea} or vice versa \cite{singh2018:fe}. 
We cast fair ranking as an unconstrained optimization problem over a ranking policy $\pi$, whose objective function is a linear combination of a utility measure and a fairness measure.
\begin{equation}
    \mathcal{Q}(\pi) = \lambda u(\rho_{\pi}) + (1-\lambda)f(R_i, A_i(\rho_{\pi})), \hspace{0.2cm \lambda \in (0,1)}
    \label{eq:obj_func}
\end{equation}
\noindent Here $u(\cdot)$ is a function computing the utility of a sequence of rankings $\rho_{\pi}$ for item consumers, produced by a policy $\pi$. For example, in an IR setting, $u(\cdot)$ is a proxy for the information gained by users from $\rho_{\pi}$. We measure the utility of a ranking (for a single impression) via normalized ERR \cite{chapelle2009:er}, where the normalization ensures that a ranking where items are perfectly sorted by relevance has utility equal to 1, regardless of the items available for ranking and their relevance. ERR is based on a cascade browsing model, where users view search results from top to bottom, with a probability of abandoning at each position which increases with rank and relevance of examined items.\footnote{Following \cite{biega2019:ot}, the probability of user stopping at position $p$, after viewing items $\left \lbrace u_1, \dots, u_p \right \rbrace$ is set to $P(\text{stop}|u_1, \dots, u_p)=\gamma^{p-1} cr_p \prod_{i=1}^{p-1}(1-cr_i), \hspace{0.2cm} c=0.7, \gamma=0.5 $} The overall utility $u(\rho_{\pi})$ is computed as the average utility over all impressions. More in general, utility can be broadly characterized as user satisfaction, potentially including notions of diversity in rankings \cite{clarke2008:nd}.

The objective function $\mathcal{Q}$ also comprises a function $f(\cdot)$ which combines the cumulative relevance $R_i$ and exposure $A_i$ of items to compute the fairness of ranking policy $\pi$ toward item providers. To this end, we follow \cite{biega2019:ot} by requiring that items receive a share of cumulative exposure that matches their share of cumulative relevance. More precisely, let us define the attention and relevance accumulated by item $u_i$ over $J$ queries as $A_i  =  \sum_{j=1}^J a_i^j$, $R_i  =  \sum_{j=1}^J r_i^j$, and let us denote as $\bar{A}_i$ and $\bar{R}_i$ their normalized versions

\gmdncomment{at this point, we're assuming that relevance $r_i^j$ may change across impressions, is it correct?} \afabcomment{yes, our notation allows for that; in the experiments, relevance is fixed.}

%\begin{align}
%A_i & =  \sum_{j=1}^J a_i^j \label{eq:A} \\
%R_i & =  \sum_{j=1}^J r_i^j \label{eq:R}
%\end{align}

%\noindent and let us denote as $\bar{A}_i$ and $\bar{R}_i$ their normalized versions

%\begin{align}
%\bar{A}_i & =  \frac{A_i}{\sum_{i=1}^I \bar{A}_i} \label{eq:A_bar} \\
%\bar{R}_i & =  \frac{R_i}{\sum_{i=1}^I \bar{R}_i}. \label{eq:R_bar}
%\end{align}

\begin{equation}
\bar{A}_i =  \frac{A_i}{\sum_{i=1}^I A_i}; \hspace{0.2cm}
\bar{R}_i =  \frac{R_i}{\sum_{i=1}^I R_i}. \label{eq:R_bar}
\end{equation}

\noindent Cumulative unfairness is then quantified by the $\ell_2$ norm of vector
\begin{equation}
    \bar{A}-\bar{R} = \left [ \bar{A}_1-\bar{R}_1, \dots, \bar{A}_I-\bar{R}_I \right ]
\end{equation}

\noindent and fairness by its negation:

\begin{equation}
    f(A, R) = - || \bar{A}-\bar{R}||_2.
    \label{eq:fair}
\end{equation}

%\afabcomment{It should be noted that normalization (Equation \ref{eq:A_bar}, \ref{eq:R_bar}) takes care of a potential problem of exploding unfairness, caused by the fact that the attention budget $\sum_{i=1}^I a_i^j$ and relevance budget $\sum_{i=1}^I r_i^j$ for a query may differ systematically, causing cumulative unfairness to grow due to systematic over- or under-exposure of items on average.}

To quantify the attention $a_i^j$ received by an item ranked in position $\rho_i^j$ we use, again, the browsing model of ERR \cite{chapelle2009:er}, so that the same user model underlies the estimates of item exposure and user utility. 

We adopt a \emph{within-query} policy for fairness amortization and focus on a single information need. Five items $(u_0, u_1, u_2, u_3, u_4)$ of relevance $r=(r_0, r_1, r_2, r_3, r_4)$ compete for attention in rankings over multiple impression of the same query. While in practice systems are required to rank large sets of items, this reduced cardinality allows us to enumerate all solutions and compute the perfect greedy solution without resorting to approximations or heuristics.  We consider three relevance distributions, corresponding to different availability of relevant items and, consequently, different targets of ideal exposure. Specifically, item relevance decreases linearly from $r_0=1$, for the most relevant item, to $r_4=\text{min}(r_i^j)$ for the least relevant one, so that $r = (1, 1-\delta, 1-2\delta, 1-3\delta, 1-4\delta)$. Three values are tested for parameter $\delta$, namely $\delta=0.25$ (large relevance difference), $\delta=0.125$ (intermediate relevance difference) and $\delta=0.05$ (small relevance difference).

%according to three different distributions: %\gascomment{generalizziamo mettendo una definizione con un 'rate' tipo $r = (1, 1-\delta, 1-2\delta, 1-3\delta, \-4\delta)$; noi consideriamo poi 3 valori di $\delta$...}
%\begin{itemize}
%    \item Large differences in relevance: $r=(1, 0.75, 0.5, 0.25, 0)$.
%    \item Small differences in relevance: $r=(1, 0.95, 0.9, 0.85, 0.8)$.
%    \item An intermediate setting with $r = (1, 0.875, 0.75, 0.625, 0.5)$. 
%\end{itemize}

%\afabcomment{Normative reasoning: what is desirable? If duplicates belong to the same entity/person collapse them, otherwise keep em separate. This does not necessarily mean that they should be treated equally: one may inherently deserve more attention due to precedence (which is arguably a proxy for ownership of written text). May want to bring forth different arguments for IR vs e-commerce and point to domain-specificity of normative reasoning.}

%\afabcomment{In some situations we may postulate that an item and its copy deserve the same attention the item would get on its own. e.g. duplicate posting on Amazon belonging to same seller. Is there any situation where a duplicate from a different seller/entity is legitimate and deserves attention? I guess a (near-) duplicate from another seeler would make sense. A duplicate in IR may be plagiarism; regardless, it seems that the postulate would hold here.}

To solve the optimization problem, we test a \emph{stateful} policy that explicitly targets the objective function (Equation \ref{eq:obj_func}) by keeping track of the relevance and exposure accumulated by items up to some impression $t$. At step $t+1$, we exploit this information to compute the best greedy solution via enumeration. This ranking policy is compared against a \emph{stateless} approach that exploits Placket-Luce (PL) sampling \cite{luce1959:ic,plackett1975:ap}. PL sampling is based on drawing the top-ranked item at random from a categorical distribution where the probability of drawing item $u_i$ is equal to $r_i^j / \sum_{i=1}^I r_i^j$. The chosen item is then removed from the pool of candidate items, from which the second ranking item is drawn in the same way, based on $\ell_1$-normalization of the relevance scores of remaining candidates. The procedure is repeated until all items are drawn. 

%To evaluate the benefits of item duplication under these ranking policies,
To evaluate the effects of item duplication in systems where copies incur different relevance penalties,
we let the cost of duplication take values $k=1$ (free copy) and $k=0.5$ (relevance of duplicate item is halved). For each combination of relevance distribution, summarized by parameter $\delta$, and cost of duplication $k$, we test each policy in six different settings. In each setting a different item $u_i$ is duplicated, and an additional setting accounts for a scenario without any duplicates.

\subsection{Results}
\label{sec:res}
%\afabcomment{Solutions on unfairness vs utility plane. Duplication incentives.}

\begin{figure}[t]
 %\captionsetup[subfigure]{labelformat=empty}
  \centering
  \begin{subfigure}{\linewidth}
    \centering
    \includegraphics[width=.95\linewidth]{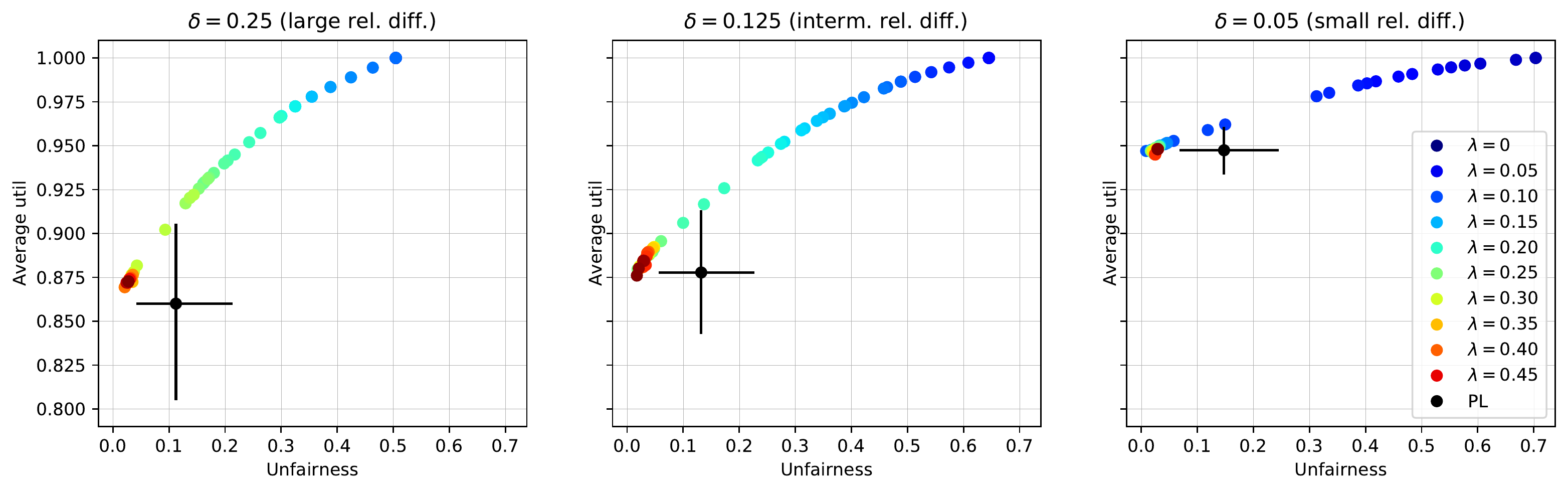}
    \vspace{-0.25cm}
    \caption{$J=20$ impressions (rare queries)}
  \end{subfigure}

  \begin{subfigure}{\linewidth}
    \centering
    \vspace{0.15cm} 
    \includegraphics[width=.95\linewidth]{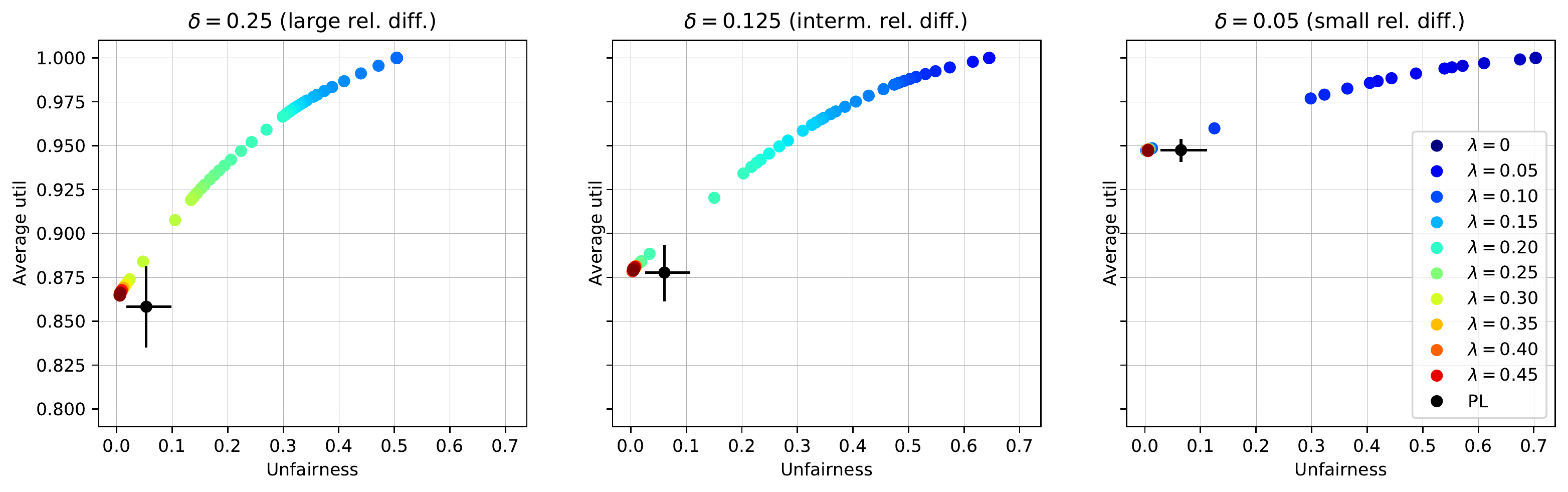}
    \vspace{-0.25cm}
    \caption{$J=100$ impressions (frequent queries)}
    \label{fig:lambda_choice_100impr}
  \end{subfigure}  
  \caption{Unfairness ($x$ axis) vs average utility ($y$ axis) for \emph{stateful} greedy solutions over different values of $\lambda$ (downsampled and color-coded in legend). In black, summary of 1,000 Plackett-Luce repetitions, reporting median, 5th and 95th percentile for utility and unfairness. Each column corresponds to a different relevance profile for the available items, namely large relevance difference (left -- $\delta=0.25$), intermediate difference (middle -- $\delta=0.125$) and small difference (right -- $\delta=0.05$). Solutions with $\lambda>0.5$ are omitted for better color-coding as they are all in a close neighbourhood of $\lambda=0.5$.}  
  \label{fig:lambda_choice}
\end{figure}

As a first step, we ensure that the stateful policy can effectively trade off relevance and fairness in the basic setting where no duplicates are present. In Figure~\ref{fig:lambda_choice}, we evaluate the impact of parameter $\lambda$ on the utility and unfairness of rankings produced by the stateful policy, where unfairness is defined as the negation of function $f(\cdot)$ in Equation \ref{eq:fair}. As a baseline, we test the PL-based policy $\pi_{\text{PL}}$, reporting median values for utility and unfairness over 1,000 repetitions, along with the 5th and 95th percentile. Each panel in Figure \ref{fig:lambda_choice} corresponds to a different combination of relevance difference, parametrized by $\delta$, and number of impressions $J$. The top row corresponds to a less frequent query ($J=20$) and the bottom row to a more frequent one ($J=100$). Panels on the left depict results for a large relevance difference ($\delta=0.25$), middle panels correspond to an intermediate relevance difference ($\delta=0.125$) and left panels to a small one ($\delta=0.05$).

We find that, over large relevance differences (left panels), a value $\lambda \geq 0.3$ is required to approach zero unfairness, while, for small relevance differences (right panels), $\lambda = 0.1$ is sufficient.  This is expected: as relevance becomes uniform across items, even a policy marginally focused on fairness ($\lambda=0.1$) can bring about major improvements in the distribution of attention. Moreover, for a small relevance difference, the trade-off between fairness and utility is less severe, which is also expected. When items have a similar relevance, a policy can more easily grant them a share of exposure proportional to their share of relevance, while only suffering a minor loss in terms of utility. Furthermore, the unfairness of exposure brought about by a solution purely based on relevance ($\lambda=0$) increases as the difference in relevance for the available items become smaller. This is a desirable property of unfairness function $u(\cdot)$. Indeed, if a small difference in relevance ($\frac{r_0}{r_4}=1.25$) corresponds to a large difference in attention ($\frac{A_0}{A_4}>1,000$), then the distribution of exposure stands out as particularly unfair for item $u_4$. 

Although dominated by the stateful approach we just analyzed, the baseline PL-based policy $\pi_{\text{PL}}$ consistently provides low-unfairness solutions. Unfairness is especially low for frequent queries, while for a rare query PL sampling is less likely to successfully distribute the cumulative exposure of items so that it matches their cumulative relevance. For frequent queries, the intervals spanned by the 5th and 95th percentile are narrower, signalling lower variance. PL sampling has been found to be a good baseline approach to obtain a fair policy from relevance scores under top-heavy browsing models such as the one underlying ERR \cite{diaz2020:es}. Overall, our experiment confirms this finding.

\begin{figure}[t]%[h!]
\includegraphics[width=\textwidth]{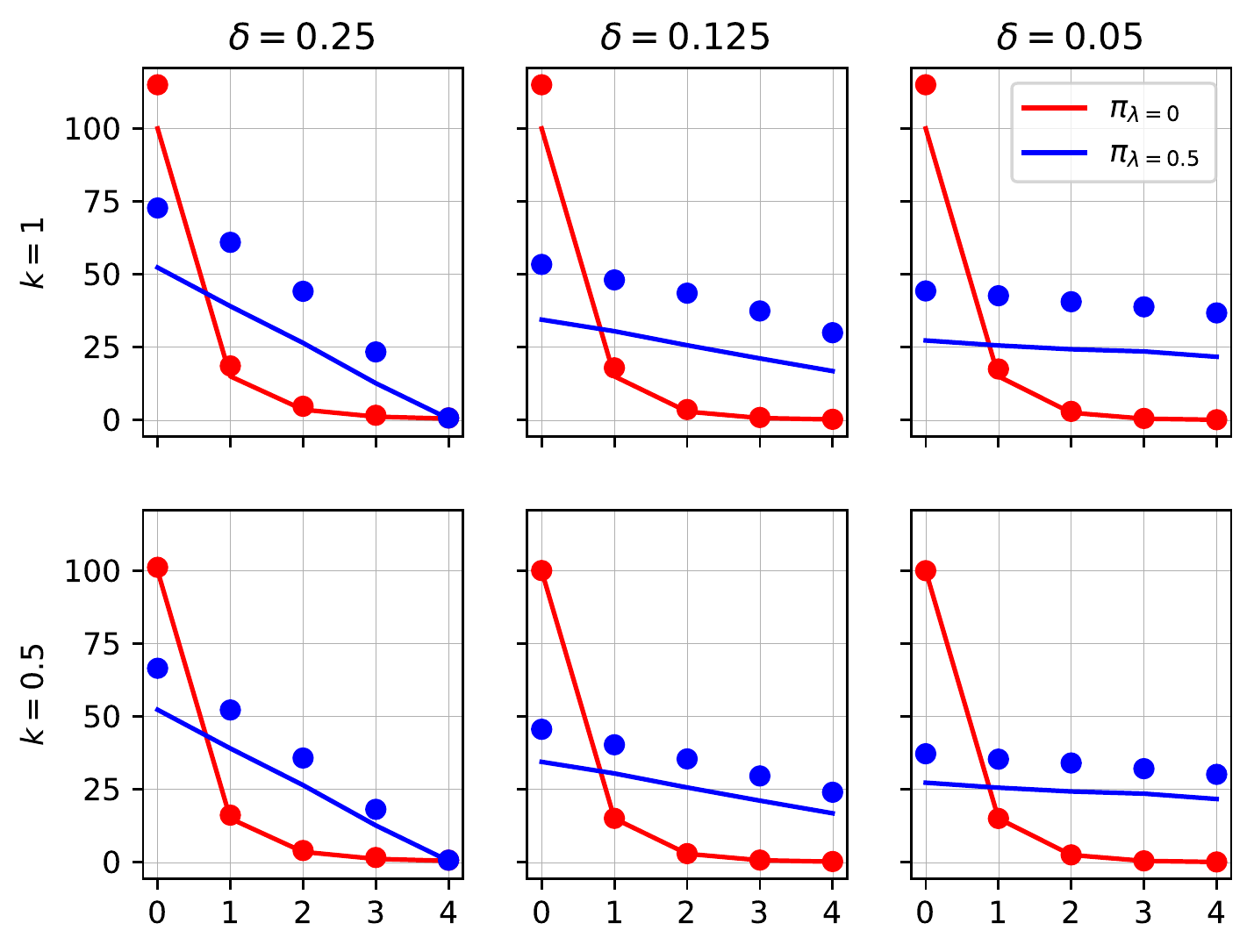}
\caption{Solid lines represent the attention $A_i$ accumulated by each item ($y$ axis) under a fairness-aware policy $\pi_{\lambda=0.5}$ (blue) or a policy solely focused on (ERR-based) utility $\pi_{\lambda=0}$ (red) in the absence of duplicates, over $J=100$ impressions of the same query. Item indices $i \in \left \lbrace 0, \dots, 4 \right \rbrace$ vary along the $x$ axis. Round markers summarize the extra-attention one item would obtain if duplicated. 
%The round marker at position $x=i$ represent the sum of the attentions $(A_i+A_{\tilde{i}})$ received by item $u_i$ and its copy $u_{\tilde{i}}$ if it was duplicated (while remaining items are not duplicated). 
Each column corresponds to a different relevance profile for the available items, namely large relevance difference (left -- $\delta=0.25$), intermediate difference (middle -- $\delta=0.125$) and small difference (right -- $\delta=0.05$). Each row corresponds to a different relevance multiplier for duplicates, namely $k=1$ (top) and $k=0.5$ (bottom).} \label{fig:duplicate_l05}
\end{figure}

To study the interaction between item duplication and equity of ranking, we firstly concentrate on a stateful greedy solution with $\lambda=0.5$, representing a policy strongly focused on fairness. In Figure \ref{fig:duplicate_l05}, solid lines represent the distribution of cumulative attention $A_i$ for each item $u_i$ under a fairness-aware policy ($\lambda=0.5$ - blue) and a policy solely focused on utility ($\lambda=0$ - red), in the absence of duplicates. The query of interest is repeated for $J=100$ impressions. Each column corresponds to a different value of $\delta$, which determines large difference (left panels), intermediate difference (middle panels) and small difference (right panels) in relevance for items $u_i$. Interestingly, cumulative attention under policy $\pi_{\lambda=0.5}$ ends up resembling the distribution of relevance $r_i^j$ for items $u_i$, i.e.\ a linear distribution with variable steepness. Policy $\pi_{\lambda=0}$, on the other hand, is not affected by the distribution of relevance.

Each round marker at position $x=i$ represents the sum of the attentions $(A_i+A_{\tilde{i}})$ received by item $u_i$ and its copy $u_{\tilde{i}}$, if said item was duplicated (while remaining items are not). In other words, compared against solid lines of the same color, round markers summarize the extra-attention one item would obtain if duplicated. Different rows in Figure \ref{fig:duplicate_l05} correspond to a different relevance multiplier for duplicates, namely $k=1$ (top) and $k=0.5$ (bottom). 

For every combination of parameters $k$ and $\delta$ considered and for each item $u_i$, duplicates are always rewarded more under a fairness-aware policy $\pi_{\lambda=0.5}$ than under a policy solely focused on relevance $\pi_{\lambda=0}$. This finding suggests that fairness in rankings may be gamed by providers who duplicate their items. Moreover, in the presence of duplicates or near-duplicates, fairness of rankings may be at odds with diversity. Duplicated items, especially top-scoring ones, end up obtaining a significant amount of extra-attention. In turn, this may incentivize item providers to duplicate their listings. If redundancy in candidate items increases, it becomes harder for a ranking system to achieve diverse rankings, with potential repercussions on user satisfaction \cite{ekstrand2014:up} and perception \cite{kay2015:ur}. As expected, however, the benefits of duplication become smaller as its cost increases (bottom panels). 

%\afabcomment{While technical solutions for near-duplicate detection and removal are certainly available, they may not always be viable, as nearly identical listings can be posted in accordance with system regulation to stress slight differences in products.}
%\afabcomment{Also, ack expl. that copies of the same item may be getting attention in different impressions.}

\begin{figure}[t]%[h!]
\centering
\begin{subfigure}{.5\textwidth}
  \centering
  \includegraphics[width=.9\linewidth]{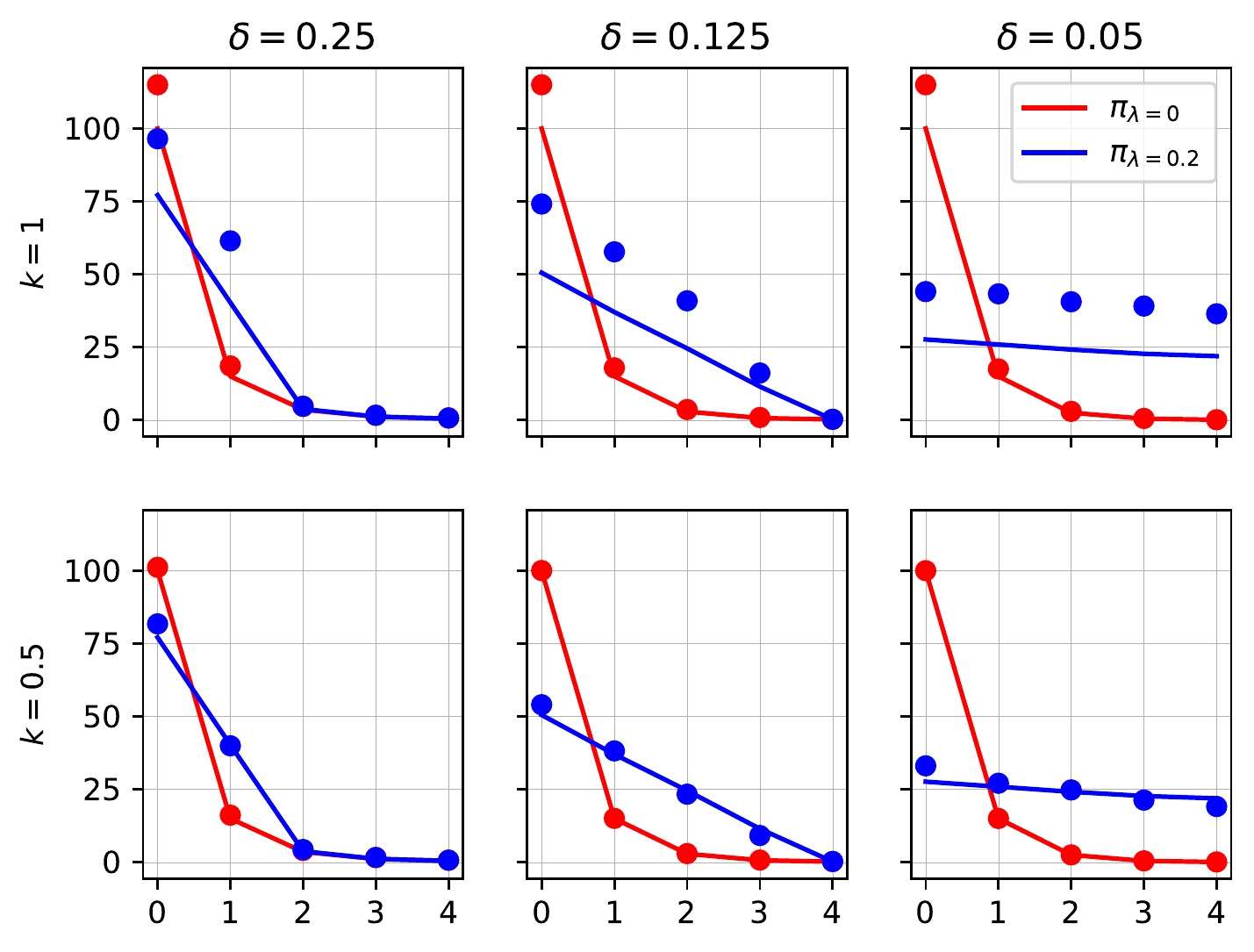}
  \caption{$\pi_{\lambda=0.2}$}
  \label{fig:duplicate_l02}
\end{subfigure}%
\begin{subfigure}{.5\textwidth}
  \centering
  \includegraphics[width=.9\linewidth]{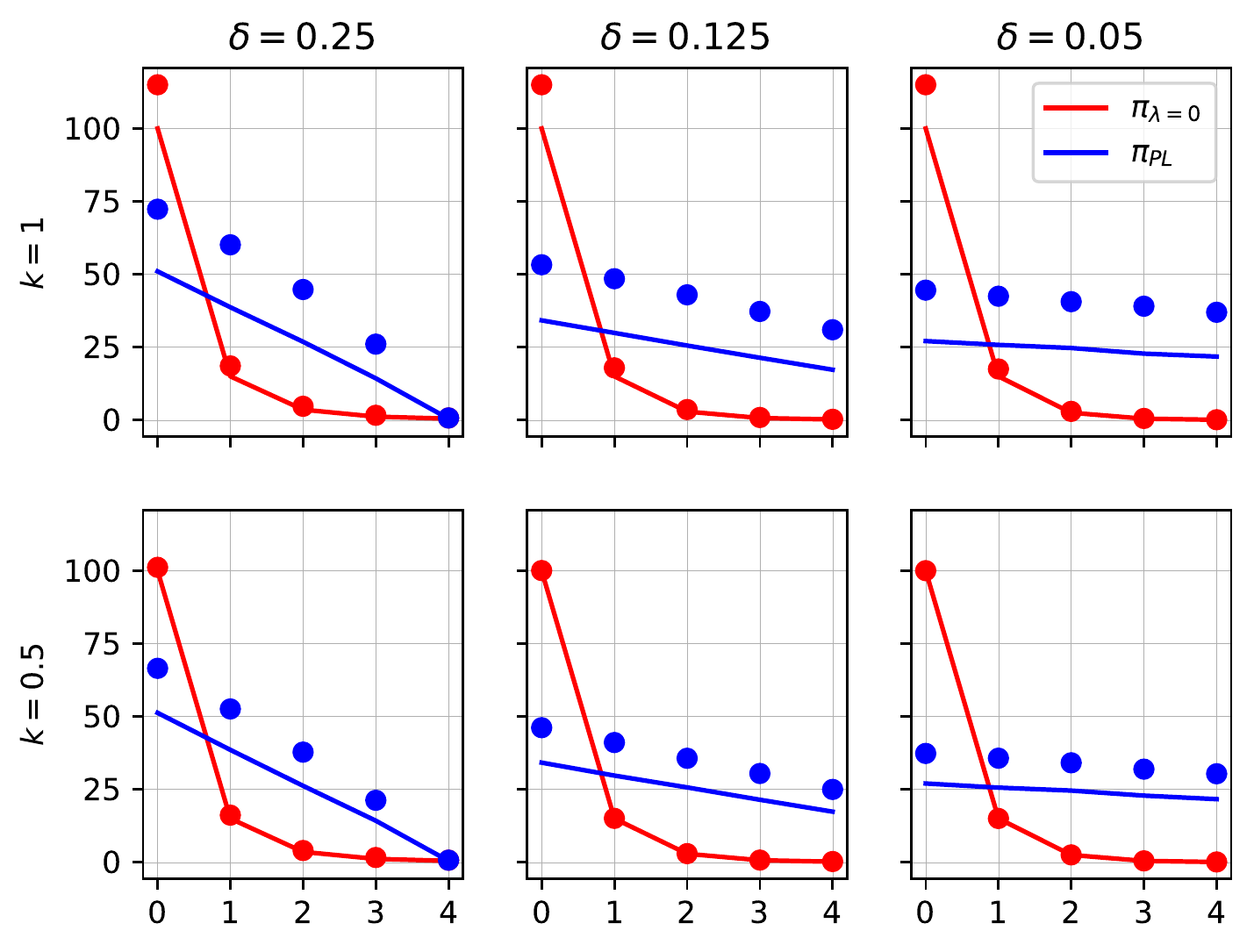}
  \caption{$\pi_{\text{PL}}$}
  \label{fig:duplicate_pl}
\end{subfigure}
\caption{Same analysis as Figure \ref{fig:duplicate_l05} with different ranking policies.}
\label{fig:duplicate_final}
\end{figure}

Figure \ref{fig:duplicate_l02} summarizes the same analysis for $\lambda=0.2$, which corresponds to a more balanced ranking policy. In general, policy $\pi_{\lambda=0.2}$ is more similar to $\pi_{\lambda=0}$, i.e.\ it is more focused on relevance and rewards duplicates less than $\pi_{\lambda=0.5}$. The most relevant items still obtain a sizeable benefit from duplication, especially when the copying process does not affect item relevance (top panels). Finally, we evaluate the extent to which a policy based on PL sampling rewards duplicates. Figure \ref{fig:duplicate_pl} reports the extra-attention obtained by duplicated items under $\pi_{\text{PL}}$. These results are similar to those obtained under policy $\pi_{\lambda=0.5}$ in Figure \ref{fig:duplicate_l05}, showing that duplicates are likely to receive a sizeable extra-exposure also under the stateless PL-based policy. This finding is not surprising given that, in order for $\pi_{\lambda=0.5}$ and $\pi_{\text{PL}}$ to achieve similarly low unfairness for frequent queries (Figure \ref{fig:lambda_choice_100impr}), they must distribute item exposure in a similar fashion.

\section{Conclusions}
\label{sec:concl}

In this work we have shown that duplicates are a potential blind spot in the normative reasoning underlying common fair ranking criteria. On one hand, fairness-aware ranking policies, both stateful and stateless, may be at odds with diversity due to their potential to incentivize duplicates more than policies solely focused on relevance. This can be an issue for system owners, as diversity of search results is often associated with user satisfaction \cite{ekstrand2014:up}. On the other hand, allowing providers who duplicate their items to benefit from  extra-exposure seems unfair for the remaining providers. Finally, system users (item consumers) may end up being exposed to redundant items in low-diversity search results; this would be especially critical in situations where items convey opinion.

While technical solutions for near-duplicate detection and removal are certainly available \cite{amazon,broder2000:if}, they may not always be viable, as nearly identical listings can be posted in accordance with system regulation, e.g.\ to stress slight differences in products. Control over near-duplicates is even weaker in web page collections indexed by search engines. Therefore, it is important to consider the entities and subjects who benefit from exposure of an item and factor them into the normative reasoning underlying a fair ranking objective. While in marketplaces beneficiaries of exposure are more easily identifiable, for document collections the situation is surely nuanced, including for instance the writer, publisher and subject of a document.

Future work should comprise a more detailed study, including cross-query measures, considering different user browsing models and richer models for duplication and its cost. Moreover, it will be interesting to systematically assess the relationship between provider-side fairness and diversity of search results in the presence of duplicates, and the extent to which these desirable objectives are in conflict with one another.

\subsubsection{Acknowledgements} We thank anonymous reviewers for their thoughtful and helpful comments.

\bibliography{references}

%Author, F.: Article title. Journal \textbf{2}(5), 99--110 (2016)

%Author, F., Author, S.: Title of a proceedings paper. In: Editor,
%F., Editor, S. (eds.) CONFERENCE 2016, LNCS, vol. 9999, pp. 1--13.
%Springer, Heidelberg (2016). \doi{10.10007/1234567890}

%Author, F., Author, S., Author, T.: Book title. 2nd edn. Publisher,

%Author, A.-B.: Contribution title. In: 9th International Proceedings
%on Proceedings, pp. 1--2. Publisher, Location (2010)

%LNCS Homepage, \url{http://www.springer.com/lncs}. Last accessed 4
%\end{thebibliography}
\end{document}